# Memory effects in permalloy-niobium hybrid structures

L.S. Uspenskaya, S.V. Egorov

*Institute of Solid State Physics RAS, 142432, Chernogolovka, Russia*

uspenska@issp.ac.ru

A.A. Chugunov

*Department of fundamentals of physical-chemical engineering MSU, Russia*

**Abstract.** The kineticts of magnetization reversal of stripe-shaped permalloy-niobium hybrid nanofilms is studied in 6-300 K temperature range by means of magneto-optics visualization technique. The niobium influence on magnetic domain walls type and on magnetic domain structure of permalloy via the interface quality and via the distortion of stray fields is found. The memory effect, which is the superconducting niobium memory about an initial magnetic domain structure of permalloy at cooling below $T_c$, is found. The memory is razed only by hybrid heating over $T_c$.

The vicinity of a superconductor (S) and a ferromagnet (F) in S/F hybrid nanostructures brings a lot of new interesting phenomena like the appearance of triplet superconductivity, "long range" proximity effects, magnetic pinning, pi-contacts, huge increase and variation of the sigh of colossal magnetoresistance [1-6]. Most of them are nowadays widely studied, but a few is known about the magnetic pinning in hybrid nanofilms and especially about magnetic hysteresis. Here we report about first observation of the hysteresis phenomena of bilayer permalloy-niobium (Py-Nb) hybrid nanostructures at magnetization reversal under the in-plane magnetic field, which we observed at T = 7 – 8 K, the temperature below the transition of niobium layer into superconducting state.

The experiments were performed on the stripe structures with stripe width about 10 to 200 microns formed by lift-off lithography on bilayer Nb-Py nanofilms grown by RF magnetron spattering successive evaporation of Nb and Py layers on cold Si substrate in the presence of an in-plane field aligned with the length of the stripes. The Nb layers were 60-100 nm thick, which provides $T_c^{Nb}$=9 K. The thickness of Py layers was varied in the range from 10 to 60 nm. The observation of magnetization reversal was performed by magnetooptic visualization technique by means of yttrium-iron garnet indicator films with in-plane magnetization [7], which provides the real time observation of kinetics of magnetization process with space resolution about 1 micron in temperature range 7 K - 300 K.

The first results concerned the variation of the type of domain walls in permalloy with niobium thickness. We found that for the Nb bottom layer as well as Nb top layer with the thickness of 60 nm the type of domain walls in Py layer was the same, as in single layer Py with corresponding thickness, Figure 1 (a,b), that is the wall were of Neel type, with magnetization vector rotation in the film plane. While in Py-Nb bilayers with 100 nm Nb layer the domain walls were of Bloch type, that is with out of plane rotation of magnetization vector, Figure 1(c,d).

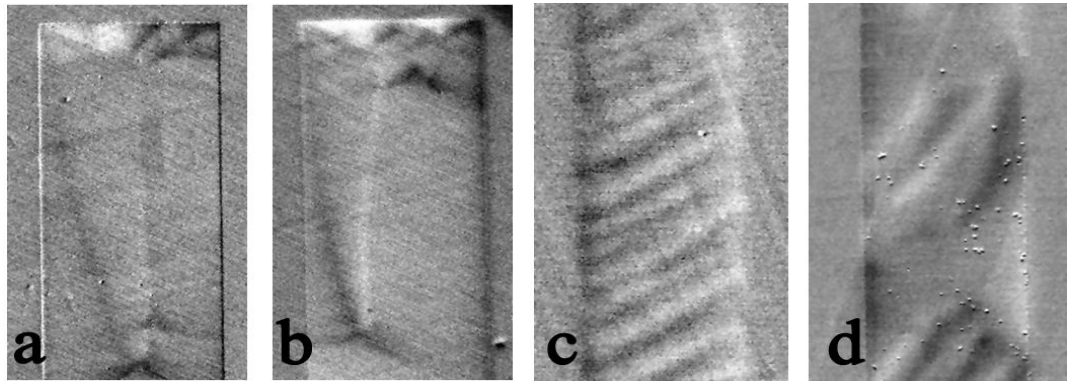

Figure 1. Domain wals in 30 nm thick Py layer with bottom and top Nb layer (a,b) and the walls in 20nm and 40 nm Py grown on 100 nm Nb layer (c,d) at room temperature. The samples in the shape of stripes are 200 micron thick and have the length of a few millimeters.

The difference was seen already at room temperature. We found that the origin in the variation of the type of the wall was in interface roughness, which increases with the bottom Nb layer thickness.

Second, the type of domain wall in Py was not changed while samples cooling down to $T_c^{Nb}$ and even below it. The magnetization reversal was also the same at $T > T_c^{Nb}$, only coercitivity of the process was increased by order of value. However, the magnetization reversal depended crucially upon the type of domain walls at $T < T_c^{Nb}$, Figure 2.

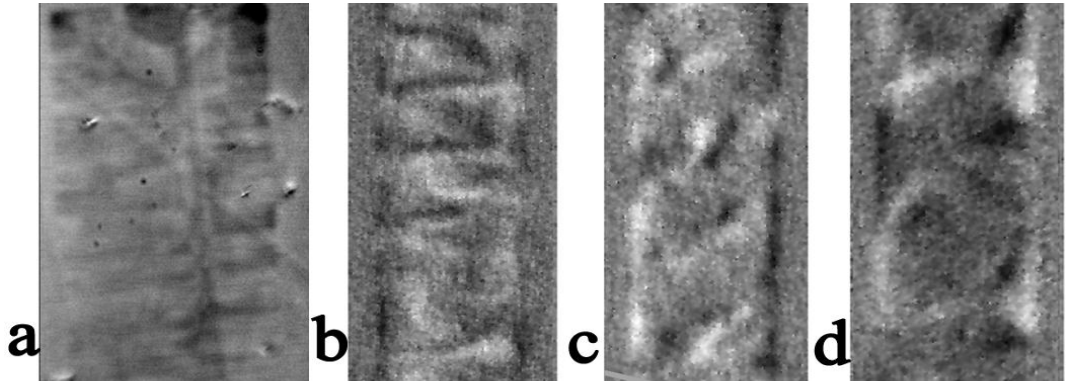

Figure 2. The magnetization reversal of the same samples, as shown in Figure 1, at T = 7 K. The width of all stripes is 200 micron.

For thin Py layer the reversal occurs via inhomogeneous magnetization rotation (a): the cross-tie closure domains on the wall transform into part-through domains (*compare Figure1(a,b) and Figure 2*), which cross the stripe from one side to another [8]. The domain walls in thicker Py layer remain the same both above and below $T_c^{Nb}$, Figure 2(b,c,d), as well as for the room temperature. However, the width of domains (or the number of the walls, which appear during magnetization reversal) becomes dependent upon the magnetic prehistory.

First, if we start cooling the bilayer while permalloy is in single domain state (SDS), then the period of domain structure is smaller, then in case of sample cooling in the presence of domain structure (PDS), compare Figures 2(b,c) and Figure 2d. Second, the domain structure period differs twice for the fields switched in opposite directions after sample cooling in SDS (on forward shoulder and on returned shoulder of magnetization loop). Third, the loop for SDS cooling is remarkably shifted relative zero field; there is the bias field of about 4 Oe (magnetization reversal started at $H_{c1}$ = 4 Oe or $H_{c2}$ = -12 Oe, for opposite directions of the field ). The last, the more surprising result, concerns the stability of the bilayer memory of the prehistory. After several cycling of the field between 0 Oe, + 200 Oe, 0 Oe, -200 Oe and so on, the sample still do not forget, that it was cooled in the presence of domains and domain structure patters is still reproduced, despite the field of 200 Oe definitely completely penetrate the Nb layer at 7 K, which is confirmed by MO observations. The coercitivity after cooling in the PDS is twice higher then after cooling in SDS.

So, we found that the type of domain walls in permalloy is very sensitive to the interface quality. The variation of the roughness from 0.1 nm to 1 nm at the layer thickness from 10 nm to 60 nm leads to the variation of the type of domain wall from the Neel type to the Bloch one.

We found that the domain wall of Bloch type are much more stable than the walls of Neel type; the walls of Bloch type are observed at magnetization reversal in wide temperature range, the Neel walls are formed only at magnetization reversal at $T > T_c^N$, while $T < T_c^N$ they are transformed into the wave of inhomogeneous magnetization rotation, which could be explained by the distortion of magnetic stray field distribution by superconducting Nb. This result evidence the validity of the suggestion about the variation of the type of magnetic domain structure in Py at $T > T_c^N$ and at $T < T_c^N$ made in [6] to explain the variation of the sigh of effect of colossal magnetoresistance during the bilayer cooling.

We show the possibility to control the magnetization reversal of bilayer Py-Nb films by variation of the domain structure of ferromagnetic layer before sample cooling. The stability of the memory effect looks surprising. We have to conclude that the vortices in Nb film induced by domain walls during cooling interact weakly with crossed vortices, which enter the film under the in-plane maghetic field even at T = 7K - 8K.